\documentclass[10pt]{article}

 \usepackage{amsmath,amsthm,amssymb}
 \usepackage{makeidx,epsfig,lscape}
 \usepackage{color,colortbl}
 \usepackage{fancyhdr}

 \definecolor{myaqua}{rgb}{0.0,0.5,0.55}
 \definecolor{lightaqua}{rgb}{0.75,0.95,0.95}

 \usepackage[colorlinks = true,
            linkcolor = myaqua,
            urlcolor  = blue,
            citecolor = myaqua]{hyperref}

\usepackage{caption}
\usepackage{floatrow}
\def\lin#1#2{\textcolor[rgb]{0.6,0.6,0.6}{\vspace*{#1mm} \hrule
   height 3 pt \vspace*{#2mm}}}
\def\bt{\begin{tabular}}
\def\et{\end{tabular}}
\def\and{\mbox{ and }}

\def\1{{\bf 1}}

 \def\sectionn#1{\refstepcounter{section}{\color{myaqua}

 \vskip 6mm

 \noindent\Large\bf\thesection. #1}

 \vskip 3mm}

\begin{document}

 \fancyhead[L]{\hspace*{-13mm}
 \bt{l}{\bf }\\
 \et}
 \fancyhead[R]{\includegraphics{pic1.ps}}

 $\mbox{ }$

 \vskip 12mm

{ 

{\noindent{\huge\bf\color{myaqua}
  The Geometro-Hydrodynamical Representation  \\[2mm] of the Torsion Field}}
%
\\[6mm]
{\large\bf Mariya Iv. Trukhanova$^1$ and Shipov Gennady$^1$}}
\\[2mm]
{ 
 $^1$Faculty of physics, Lomonosov Moscow State University, Moscow, Russian Federation\\
Email: \href{mar-tiv@yandex.ru}{\color{blue}{\underline{\smash{mar-tiv@yandex.ru}}}}\\[1mm]
\\

\lin{5}{7}

 { 
 {\noindent{\large\bf\color{myaqua} Abstract}{\bf \\[3mm]
 \textup{    We construct the geometro-hydrodynamical formalism for a spinning particle based on the   six-dimensional manifold of autoparallelism geometry which is represented as a vector bundle with a base formed by the manifold of the translational coordinates and a fibre specified at each point by the field of an orthogonal coordinate frame underlying the classical spin. We show that the geometry of oriented points leads to the existence of torsion field with the source - the classical spin. We expand   the geometro-hydrodynamical representation of Pauli field developed by Takabayasi and Vigier.    We show that the external torsion field has a force effect on the velocity and spin fields via the spin-vorticity, which is characteristic of the space structure with the inhomogene triad field.  The possible experimental effects of torsion field are discussed.
 }}}
 \\[4mm]
 {\noindent{\large\bf\color{myaqua} Keywords}{\bf \\[3mm]
               torsion, spin, vorticity
}

 \fancyfoot[L]{{\noindent{\color{myaqua}{\bf }} }}

\lin{3}{1}

\sectionn{Introduction}

{ \fontfamily{times}\selectfont
 \noindent

The main concept of quantum-mechanical theory of a single non-relativistic particle with spin is a spinor wave field. The construction  of a realistic physical model of a quantum matter on the way of representing visually the information encoded in a spinor was undertaken in Refs.  \cite{1} - \cite{5}. The new  geometrical representation of a two-component spinor with an orthogonal set of coordinate axes was developed by Kramer \cite{1}. The representation of Pauli field as an assembly of very small  rotating particles continuously distributed in space was later evolved in Ref. \cite{2} using Kramers' idea. The hydrodynamical consideration describes the Pauli field as a fluid of small particles having intrinsic angular momentum or spin. The geometrical representation introduces a triad structure underlying the classical spin \cite{3}    where each fluid element is represented by this triad and spin vector is fixed to the third axis of the triad. Based on this, the spinor is interpreted as defining a state of rotation \cite{6}.   This approach  leads to a  concept of the spinor $\psi$ in terms of Euler angles which determine the orientation of the triad with respect to the fixed set of Cartesian axes \cite{3}.  The realistic geometro-hydrodynamical interpretation of the quantum matter is based on the postulate that embedded in the wave is a particle of mass m, which transfers energy and momentum. The evolution of the spinor field, which must represent a new physical field, implies the rotation of the triad and, as a result, the movement of the center of mass of the particle.  The transformation and orientation of the ensemble of spinning particles were described in the field variables (basic quantities $\rho$, $\mathbf{s}$, $\mathbf{v}$)  on the basis of the  conservation equation for the spatial distribution of this ensemble   \cite{4}, \cite{6} $\rho(\mathbf{r},t)=\psi^+\psi$

\begin{equation} \label{n}
\partial_t \rho +\nabla( \rho\mathbf{v})=0, \end{equation}
of the  velocity field $\mathbf{v}(\mathbf{r},t)$ evolution equation or Euler equation of flow \cite{6}

    \begin{equation} \label{j} m(\partial_{t}+\mathbf{v}\mathbf{\nabla})\mathbf{v}=
e\mathbf{E}+\frac{e}{c}\mathbf{v}\times\mathbf{B}
-
\nabla(\Pi+\Pi_{in})+\frac{e}{mc}s_k\nabla (B^k+B^k_{in}),\end{equation}
                         and of the spin field $\mathbf{s}(\mathbf{r},t)$ equation of motion \cite{6}
                   \begin{equation} \label{s}      (\partial_{t}+\mathbf{v}\mathbf{\nabla})\mathbf{s}=\frac{2\mu}{\hbar}\mathbf{s}\times(\mathbf{B}+\mathbf{B}_{in}). \end{equation}
 In this hydrodynamical representation the motion of non-relativistic quantum-mechanical spinning particle is carried out in such a way that each element of the fluid evolves like a classic spinning particle under the action of external forces and of the internal potential $\Pi_{in}=-|\nabla\mathbf{s}|^2/2m$ and  the internal magnetic field    $\mathbf{B}_{in}=c/e(\triangle\mathbf{s}+\partial_k\rho\cdot\partial_k\mathbf{s}/\rho)$.

   \section{Geometrical representation}
     We use the geometro-hydrodynamical scheme that had been evolved in the works of Takabayasi, Bohm, Vigier and Holland \cite{1} - \cite{6} and consider the Pauli field as a fluid of spinning elements.      In this section we introduce the internal configurational variables that are attributed to each fluid element. To take into account the rotational degrees of freedom in non-relativistic case we have to use the three-dimensional manifold of oriented points.  The simplest generalization of the three-dimensional Euclidian  geometry to the case of the manifold with all its points oriented is the geometry of absolute parallelism (autoparallelism $A_3(3)$) constructed on the six-dimensional manifold which is represented as a vector bundle with a base formed by the manifold of the translational coordinates and a fibre specified at each point by the field of an orthogonal coordinate frame or triad $\mathbf{e}_{(a)}(\mathbf{x},t)$, which vary from point to point or depending on $\mathbf{x}$, where $a$ is the number of the reference vector \cite{60} - \cite{62}.  The autoparallelism was used by Albert Einstein in his attempts to construct a classical unified field theory \cite{7}, which was based on the Riemann metric and autoparallelism. He had developed this geometry in the $n$-dimensional differentiable manifold.    The geometry with absolute parallelism was first considered in the works of Weitzenbock and Vitali \cite{8}, \cite{9}. Weitzenbock suggested the existence of the $n$-dimensional manifold with coordinates of Riemannian spaces with a zero curvature tensor.

     The orientation of the triad with respect to the fixed set of space axes  is defined by the space-dependent set of Euler angles $\phi(\mathbf{x},t), \vartheta(\mathbf{x},t), \chi(\mathbf{x},t).$   The system of vectors $\mathbf{e}_{(a)}$, which are defined at each point of space,  is orthonormal, determines a state of rotation $e^{(a)}_i\cdot e^j_{(a)}=\delta_i^j$ and exists as basic vectors defined  and translatable in the absolute sense to any point of space in any direction. At every world point, the orientation of the body frame can
be specified referring to the triad the origin of which coincides with the center
of mass of the particle at that moment \cite{91}.  Using the approach developed in \cite{5}, \cite{6}, we introduce the matrix  transformation  $a_{ij}$ denoting the components of triad $\mathbf{e}_i^{(a)}$ with respect to the space axes in standard orientation $\mathbf{e}_i=a_{ij}\mathbf{c}^j$. The tensor of angular velocity of the rotation of the reference frame $\omega^{j}_{k}$ can be determined by the relation
                                  \begin{equation} \label{w}  \omega_{jk}=\mathbf{e}_k\cdot\frac{d\mathbf{e}_j}{dt}, \qquad \omega_{jk}=-\omega_{kj}.
                                                            \end{equation}

                                                            Now it is natural to introduce the main properties of absolute parallelism geometry (autoparallelism). The main consequence
 of this geometry that the torsion is a characteristic of the space.  The concept of connection of geometry can be written as  $\Delta^i_{jk}=\Gamma^i_{jk}+\gamma^i_{jk},$ where the Christoffel symbols $\Gamma^i_{jk}=1/2 g^{im}(\partial_{k}g_{jm}+\partial_{j}g_{km}-\partial_{m}g_{jk}), $ which depend on the metric tensor $g_{ij}=e^{(a)}_ie^{(a)}_j$ and  $\gamma^i_{jk}$ is the Ricci rotation coefficients or torsion. The Christoffel symbols are the particular case of the affine connection and in the flat space must vanish. But in the case of the flat space with torsion the geometry is not completely described by the metric only but the independent characteristic - torsion. The general form of the torsion tensor \cite{61}, \cite{62} and \cite{10} in the coordinate indices is

 \begin{equation}   \gamma^i_{jk}=-\Omega^i_{jk}+g^{im}(g_{jn}\Omega^n_{mk}+g_{kn}\Omega^n_{mj}),
                                                            \end{equation}
                                                             where $\Omega^n_{mk}$ has the form of
                                    \begin{equation} \label{W}  \Omega^i_{jk}=\frac{1}{2}e^i_{(a)}\biggl(\partial_je^{(a)}_k-\partial_ke^{(a)}_j\biggr),
                                                            \end{equation}
                              and  can be characterized as  the object of anholonomity \cite{60}, \cite{61}, \cite{10}. In the case of the flat affine space with torsion, where the quantities $\gamma^i_{jk}$ represent the local spin connection of  space and are referred to as the Ricci rotation coefficients for the basis $e^{(a)}_i$. Torsion is an independent characteristic of the space-time and in the anholonomic coordinates the Ricci rotation coefficients are transformed as follows $\gamma^{(a)}_{(b)k}\equiv e^{(a)}_i\gamma^i_{jk} e^j_{(b)}$.
                                A space of events has two metrics - the Riemann flat metric and the three-dimensional Killing-Cartan metric $d\nu^2=d\chi_{ij}d\chi^{ij},$ where the infinitesimal increments can be given by the vector $d\chi^i=\omega^idt$.

                                                            The  parallel displacement of the triad relative to the connection $\Delta^i_{jk}$ equals zero identically $\partial_ke^{(a)}_j-\Delta^i_{jk}e^{(a)}_i=0$. From this definition the connection can be defined as $\Delta^i_{jk}=e^i_{(a)}\partial_ke^{(a)}_j$ and the Ricci rotation coefficients proportional to the covariant derivative with respect to the Christoffel symbols
                                                            $$\gamma^i_{jk}=e^i_{(a)}\nabla_k e^{(a)}_j,$$ as a result the angular velocity of the triad must have the geometrical form
                                                       \begin{equation} \label{New} \omega^i_j=\gamma^i_{jk}\frac{dx^k}{dt}.
                                                                      \end{equation}

                                                             The angular velocity $3$-vector $\omega^i$ can be found from the relations defining rigidly rotating
Cartesian coordinates and easy  to derive that the vector of the triad, which provide a covariant specification of a state of rotation, in the flat space satisfy the equation
                                                         \begin{equation}  \label{e} \frac{de^{(a)}_j}{dt}+\gamma^i_{jk}\frac{dx^k}{dt}e^{(a)}_i=0,
                                                                         \end{equation}
                                                             which is responsible for the temporal dynamics of the triad vectors. Note that even in zero external magnetic field,
                                                             when the particle has no magnetic momentum, the vectors of the triad   will precess due to the action of the torsion torque, which
                                                             will be derived below.
                                                                We resort to  the description in the set of variables $\rho$,  $\mathbf{e}_{(a)}$ and $\mathbf{v}$.
                                                           The main dynamical property of the triad is that its angular momentum of rotation or spin vector is fixed to the third axis of triad and has the magnitude $\hbar/2$
                                                           \begin{equation}  \mathbf{s}=\frac{\hbar}{2}\mathbf{\Sigma}, \qquad where \qquad   \mathbf{\Sigma}=\mathbf{e}^{(3)}=\mathbf{e}^{(1)}\times\mathbf{e}^{(2)}.
                                                                         \end{equation}

In the terms of Euler angles, the polarization - is the proper internal degrees of freedom,
can be represented as $\Sigma_1=\sin\vartheta\cos\phi$,  $\Sigma_2=\sin\vartheta\sin\phi$ and $\Sigma_3=\cos\vartheta$,
where angles $\vartheta$, $\phi$ denote the polar angles of polarization, but $\chi$ characterizes the rotational
orientation of the orthogonal vectors $\mathbf{e}^{(1)}$ and $\mathbf{e}^{(2)}$ axes on the plane normal to polarization vector.

 %


\subsection{The Pauli equation}
 Let's move on to the relativistic task and consider a four-dimensional manifold such that at each point of space the tetrad is specified.  One introduces the field of the tetrad $e^a_{\beta}$, where Greek indices are the coordinate indices  and refer to the generic inertial frame.   Tetrad defines the fundamental metric tensor $g_{\alpha\beta}=e^a_{\alpha}e^b_{\beta}\eta_{ab}$, where $\eta_{ab}$ - is the Minkowski  metric. We also discount, that the covariant derivative in nonholonomic coordinates has the form $\nabla_aF^b=\partial_aF^b+\Delta^b_{ac}F^c.$  The wave equation of the spinning particle exhibit the properties that relates to an general requirements space-time symmetry. The spinor wave function $\Psi$ for a spin-$1/2$ particle of rest mass $m$ and charge $e$ in  space-time with torsion  obeys the Dirac equation in Cartesian coordinates

                           \begin{equation} \label{Dirac}   \gamma^{\mu}D_{\mu}\Psi+\frac{imc}{\hbar}\Psi=0,
                                              \end{equation}  where we obtained the sought equation  by replacing  the operator of $4$-momentum and  introduce the covariant derivative of a spinor. The affine connection $\Delta^{\alpha}_{\beta\mu}$ different from Christoffel symbol means that the geometry is not completely
described by the metric, but has another, absolutely independent characteristic - the tensor  \cite{12}. The covariant derivative can be formed from the sum of partial derivative and the additional term - torsion
           \begin{equation} \label{covariant}   D_{\mu}=\partial_{\mu}-\frac{q}{c}A_{\mu}-\gamma_{\mu},
                                              \end{equation}
                                                             here   the nonholonomic  spin connection $\gamma_{\mu}$    represents the connection of the flat space-time but having the torsion. Space-time with zero curvature is also referred to as a Weitzenbock space-time and in that case connection can be determined via the Ricci rotation coefficients $\gamma_{\mu}=1/4\beta_a\beta_b\gamma^{ab}_{\mu},$ where the transformation of indices be carried out with the tetrad  $\gamma^b_{\mu m}=e^b_{\alpha}\gamma^{\alpha}_{\mu\nu}e^{\nu}_m.$ Via the standard conversion the spin connection obtain the form
                                 \begin{equation} \label{connection} \gamma_{\mu}=-\frac{1}{2}\sigma_{ab}\gamma^{ab}_{\mu},
                                              \end{equation}                 where the matrix $4-tensor$  for spin-$1/2$ has the representation from the Dirac matrixes $\sigma_{ab}=1/2(\beta_a\beta_b-\beta_b\beta_a).$
                                           We take into account the representation a quantity $\sigma_{ab}=(\mathbf{a},i\mathbf{\Sigma})$    and the Clifford algebra in the space with zero curvature satisfies  the permutation relation $\beta_a\beta_b+\beta_b\beta_a=2g_{ab}$

                                        \begin{equation}
                                              \beta^0=\begin{pmatrix} I & 0 \\ 0 & -I \end{pmatrix}, \qquad  \alpha^{i}=\begin{pmatrix} 0 & \sigma^{i} \\ \sigma^{i} & 0 \end{pmatrix}, \qquad \alpha^{i}=\beta^0\beta^{i}, \qquad i,j =  1, 2, 3  \end{equation} where the internal degrees of freedom corresponding to spin, or spin matrix

                                                    \begin{equation}
                                             \Sigma^{j}=\begin{pmatrix} \sigma^{j} & 0 \\ 0 & \sigma^{j} \end{pmatrix}.            \end{equation}
                                             The representation of the Pauli matrices has the form \begin{equation}
                                             \sigma_1=\begin{pmatrix} 0 & 1 \\ 1 & 0 \end{pmatrix}, \qquad  \sigma_2=\begin{pmatrix} 0 & -i \\ i & 0 \end{pmatrix}, \qquad \sigma_3=\begin{pmatrix} 1 & 0 \\ 0 & -1 \end{pmatrix}    \end{equation}

                                               Lets apply the standard procedure where for the two-component bispinor $\Psi=\begin{pmatrix}\psi \\ \varsigma\end{pmatrix}exp(-\frac{imc^2t}{\hbar})$ at low  speeds the relation is valid $\varsigma << \psi.$    In this approximation  the one-particle Pauli-Schrodinger equation for the spin $1/2$
particles motion in an external fields  has the form
                                                  \begin{equation}   \label{H}
            i\hbar\frac{\partial\psi}{\partial t}=\hat{H}\psi,
  \end{equation}
                       with the Hamiltonian operator of quantum particle

                                 \begin{equation}  \label{Hh}
                              \hat{H}=\frac{1}{2m}\Biggl(\hat{p}_{i}-\frac{q}{c}A_{i}-\frac{\hbar}{2}\gamma_{i}\Biggr)^2
                              +qA_{0}+\frac{\hbar}{2}\gamma_0-\mu\hat{\sigma}_{i}B^{i}_{ext}-\frac{\hbar}{2}\hat{\sigma}_{i}T^{i}
  \end{equation}
where     $\textbf{A}_{ext}, A_{0} $  -  are   the vector and scalar potentials of
external electromagnetic field, $\hat{\sigma}_{i}$ are the Pauli matrices, $\mu=g\mu_B/2$, $\mu$ - is the electron  magnetic moment
and $\mu_{B}=q\hbar/2mc$  is the Bohr magneton,  $q$ stands for the charge of electrons $q_e=-e$ or for the charge of ions $q_p=e$, and $\hbar$ is
the Planck constant, $g\simeq 2.0023193$, $m$ denotes the mass  of  particles, c
 is the speed of light in vacuum and the torsion has the components $\gamma_{\mu}=(\gamma_0,\gamma_j).$    In our geometrical representation triad, as the element of space, has rotational degrees of freedom and determines the fibre specified at each point of the manifold, where  the torsion field component $\gamma_{k}$ determines the rotational movement of the fluid (in the hydrodynamical representation)

 \begin{equation}
                    \mathbf{\gamma}_k= \gamma^{(1)(2)}_{k}=\partial_{k}e^{(1)}_i\cdot e^{(2)}_i.
  \end{equation}            The Hamiltonian contains additional term which is called the spin-vortex coupling $\mathbf{\Sigma}\cdot\mathbf{T}$ and can be interpreted as coupling between spin and vorticity. The main feature of a spinning hydrodynamical field is the spin-vorticity $\mathbf{T}$, which plays an important role and  characterizes the orbital motion as a result of the spatial variation of the triads  \cite{5}

 \begin{equation}  \label{vort}
                     T^{k}=\mathbf{T}=\frac{\hbar}{2m}\nabla\times\mathbf{\gamma}^{(1)(2)}=\frac{\hbar}{2m}\varepsilon^{kmn}\partial_{m}e^{(1)}_i\cdot \partial_{n}e^{(2)}_i, \qquad and \qquad div \mathbf{T}=0.
  \end{equation}

  Spin-vorticity vector has now the geometrical structure, being the curl of the torsion field component. The Hamiltonian (\ref{Hh}) denotes on the physical effects of the torsion that reminds of the vector potential influence, where the quantities $\mathbf{T}$ similarly of the magnetic field. We developed the formulation of Pauli theory combine it with the hydrodynamical and geometrical methods, where the Ricci rotation coefficients are the potential of the torsion field, which characterized by the variation from point to point of the field of frames $\mathbf{e}^{(a)}$. The influence of the torsion field on the spinning free particle is carried out through the hydrodynamic quantity - the vorticity vector $\mathbf{T}$. The orientation of the triad field is determined by the set of Euler angles and,  in  flat-space with zero curvature, but with torsion the triad field has the form

                                \begin{equation}   \label{triad}
                                              e^{(a)}_i=\begin{pmatrix}   \cos\phi\cos\vartheta\cos\chi-\sin\phi\sin\chi & \sin\phi\cos\vartheta\cos\chi+\cos\phi\sin\chi & -\sin\vartheta\cos\chi \\  -\cos\phi\cos\vartheta\sin\chi-\sin\phi\cos\chi & -\sin\phi\cos\vartheta\sin\chi+\cos\phi\cos\chi & \sin\vartheta\sin\chi \\  \cos\phi\sin\vartheta & \sin\phi\sin\vartheta & \cos\vartheta \end{pmatrix},   \end{equation} with $0\leq\phi\leq 2\pi,$  $0\leq\vartheta\leq \pi$, $0\leq\chi\leq 2\pi$, being the invariance with respect to rotations around their respective symmetry axes,   and the representation spin-vorticity vector and momentum vector in terms of Euler variables $\phi, \vartheta, \chi$ have the form of   $\mathbf{T}=\hbar/2\sin\vartheta\nabla\vartheta\times\nabla\phi$ and $\mathbf{\gamma}^{(1)(2)}=-(\nabla\chi+\cos\vartheta\nabla\phi)$. Using the definition (\ref{triad}) and the condition that the spin vector of the particle is collinear with one of the triad axes, we can represent the spinor field by the set of variables  ($\rho, \phi, \vartheta, \chi$).

     \section{The hydrodynamics equations for the spinning 1/2 particle in an external torsion field}
                                         In this section we derive the one-particle quantum hydrodynamics equations from one-particle Pauli equation (\ref{H}). We obtain equations for the charged particle with spin $1/2$.  This theory can equivalently be comprehended as a hydrodynamics with intrinsic angular momentum which creates the torsion.  The Pauli field $\psi$ can be represented by the set of hydrodynamical quantities, in the other words   the hydrodynamics description is based on the set of  vector quantities  - the vorticity  $\mathbf{T}(\mathbf{x},t),$ a vector velocity field $\mathbf{v}(\mathbf{x},t)$  and an axial vector spin field $\mathbf{s}(\mathbf{x},t)$  associated with the ensemble of oriented points.
                                          Applying of the Pauli equation with Hamiltonian (\ref{Hh}) leads to 
                            the velocity field of particle representation  in the form 


\begin{equation}\label{v}   m\mathbf{v}=\mathbf{p}-\frac{q}{c}\mathbf{A}+\frac{\hbar}{4}\sum_{a}\varepsilon^{ijk}\nabla  e_{j}^{(a)}\cdot e_{i}^{(a)}e^{(3)}_{k}, \end{equation}
where the last term represent the part of velocity field via the rotational orientations of the continuously distributed triads, it is assumed that the torsion field dynamically evolves so the triad rotates and the center of mass moves.   The last term in the velocity field of our theory (\ref{v}) coincides with  the velocity field in the hydrodynamical representation reviewed in \cite{4}, \cite{6}, but realizes  a geometrical structure.

The  momentum balance equation can
be obtained in the form

\begin{equation}\label{j3} m(\partial_{t}+\mathbf{v}\mathbf{\nabla})\mathbf{v}=
q\mathbf{E}+\frac{q}{c}\mathbf{v}\times\mathbf{B}
-
\frac{\nabla_{j}\mathbf{p}^{j}}{\rho}
+\mu  s_{j}\mathbf{\nabla}B^{j}+\mathbf{\Lambda}_{torsion},
\end{equation}
where  the torsion force field is \begin{equation}\label{j4}\Lambda^i_{torsion}=(\mathbf{v}\times\mathbf{T})^i+\frac{s_{j}}{m}\mathbf{\nabla}^iT^{j}-\frac{\partial\gamma^i }{\partial t}-\nabla^i\gamma_0.
                                                                                                 \end{equation}

  The first two terms at the right side of the equation (\ref{j3}) describe the interaction with external electromagnetic field, where the first term represents the effect of external electric field on the charge density and the second term represents the Lorentz force.    The fourth term is the effect of non-uniform magnetic field on the magnetic moment.    The fourth term appears in the equation of motion (\ref{j3}) through the magnetization energy and represents the Stern-Gerlach force due to the coupling between magnetic moment and magnetic field.  The torsion force field represented by the last term $\mathbf{\Lambda}_{torsion}$. The first term in the definition (\ref{j4}) is the torsion-dependent term and characterizes the effect of spin-vorticity on the moving particle.  The second term at the right side represents the spin-torsion coupling in the non-uniform spin-vorticity field. The last term contains the Ricci rotation coefficients and generates the influence on the particle density analogues the effect of electric field on the charge density.

  The  new torsion terms   have the new structure and the model developed here demonstrates the connection between geometry with torsion and geometro-hydrodynamical formalism developed by Takabayasi \cite{2} - \cite{6}. In the geometro-hydrodynamical representation the torsion field is associated now with the fluid constituted of a continuous distribution triads defined in the realistic hydrodynamical representation.

The equation of motion for spin vector of particle $\mathbf{s}$ in the space with torsion can be derived using the definition for the axial vector of spin density
 $\psi^+\sigma^{\alpha}\psi$

\begin{equation}  \label{M3}(\partial_{t}+\mathbf{v}\mathbf{\nabla})\mathbf{s}=\frac{2\mu}{\hbar}\mathbf{s}\times\mathbf{B}
+\mathbf{s}\times\mathbf{T},
         \end{equation}
                     where  the first term at the
right side of the equation (\ref{M3})  represents the torque caused by the interaction with the external magnetic field. The second term is the spin-torsion coupling term  and can
be interpreted as the torque caused by the interaction with spin-vorticity $\mathbf{T}$.

    To close the  equations set (\ref{n}) and  (\ref{M3}) we need to derive
the equation for the spin-vorticity   evolution, which can be obtained in the general form

      \begin{equation}\label{T}   \frac{\partial\mathbf{T}}{\partial t}=\nabla\times(\mathbf{v}\times\mathbf{T})+\frac{q}{m^2c}\nabla s_j\times\nabla B^j+\frac{1}{m}\nabla s_j\times\nabla T^j, \end{equation}  which represents the generalized vorticity equation for the spinning particle in the external torsion field.   

\subsection{The concept of torsion field}

The main concept of the new geometro-hydrodynamical formalism is its being based on the six-dimensional space of the oriented points and the orthogonal coordinate systems $e_{k}^{(a)}$ or triads are attached  in each point of space. The method depicts the torsion field with rotational motion of an assembly of triads, where we represent the torsion field in explicit geometric terms through the Ricci rotation coefficients $\gamma^{\alpha}$, on the one hand, and  introduce the new formalism of the torsion field that has now the realistic hydrodynamical representation. Using this we interpret evolution of torsion field through the field of triads, which vary from point to point via the set of Euler angles. As a result, if the torsion field dynamically evolves, the axes rotate in a corresponding fashion and the geometry modifies. Because the spin vector of the particle is collinear with one of the triad axis $\mathbf{e}_3$, torsion field must directly influence the polarization vector or spin of a particle. The method is unified with the hydrodynamical representation, in which the torsion field affects the neutral particle via the spin-vorticity which exists as a result of inhomogeneity of the triad field.

In the regions of electromagnetic fields absence the torsion field can influence the spinning particle evolution via the phase of wave function. The "canonical" momentum (\ref{v}) contains the torsion term that leads to the effect of torsion $\gamma^k$ on a neutral particle wave function.  To distinguish effects due to spin-vorticity vector $\mathbf{T}$ from effects due to torsion, a region of space should be experimentally arranged, where $\mathbf{T}=\nabla\times\gamma=0$, whereas $\gamma^k\neq 0$, because the phase of a particle wave function is affected as it passes through the $\gamma^k\neq 0$ region. The hydrodynamic  interaction is achieved by promoting the canonical momentum from $\hat{\mathbf{p}}=-i\hbar\hat{\nabla}$, to a new momentum  $-i\hbar\hat{\nabla}^k-\frac{\hbar}{2}\gamma^k.$ We use complex wave function and measurements  tell  us about the magnitude, as a result, the information encoded by phase is lost.
The external torsion field dynamically evolved through the field of frames $e_{k}^{(a)}$, which rotate in the corresponding fashion and the spinning particle is feeling the torsion field, which affects the center of mass and the spin vector. On the other hand, the spin of particle must create its own torsion field. We know that  the mass determines the affine connection of Riemannian space and creates the gravitation field, the charge creates the electromagnetic field. In the new paradigm developed in this article the fundamental property of matter is that intrinsic angular momentum  or spin, characterized by the triad,  must create the new torsion field and appears the source of torsion which  is given by the third rank Ricci tensor $\gamma^i_{jk}$.

\subsection{The experimental investigations of the torsion field}
The possible experimental observation of the torsion was discussed in Refs \cite{Ex} - \cite{Ex4}.
As we can see using generalized equations systems (\ref{j3}) and (\ref{M3}), the torsion-dependent terms bear the analogy to the terms that contain the magnetic field or vector potential-dependent terms. The main distinction is that the torsion field has different physical nature, which was discussed above.  The first experimental possibility of the torsion field influence detection is the observation of Aharonov-Bohm-like effects for the spinning neutral matter. Our theory predicts that when a beam of neutral spinning particles passes through a region of space in which there is no electromagnetic field, but there is torsion field, an interference pattern can change. The torsion field must be generated by systems of spinning particles and it appears that the torsion field inside the system must have some kind of non-local effect on the particles. On the other hand, in fact, the torsion influence is a purely local interaction with the torsion vector potential $\gamma^k$. The torsion potential is defined by connectivity on a principal bundle of the autoparallelism geometry. The changes in the fibre bundle space can lead to the phase shift on wave-function of the particles.
The second way of torsion registration is the spectral analysis experiments. As we can see from the equation of motion (\ref{j3}) the influence of torsion field can lead to splitting and shifting of known spectral lines. The spin evolution equation contains the new term that characterizes the torsion-torque generated by spin-vorticity coupling $\mathbf{s}\times\mathbf{T}.$ Torsion torque can lead to the precession of spin in the external torsion field. The equation of motion (\ref{j3}) predicts   the existence of spiral or circular trajectories of neutral particles in external torsion fields.
It is important that torsion fields  attract increasing interest  \cite{11} - \cite{15}.  The main interest lies in the direction of possibility of the experimental observation of the splitting of the known spectral lines \cite{12}, using the spectral analysis
experiments. Such experimental possibility was discussed for the splitting of the spectral lines for the hydrogen atom \cite{Ex1}.    The main target of many articles is to find the way of generation and detection of a torsion wave. It is supposed that the torsion wave can be created by coherent spin flipping a large number of particles with intrinsic spin \cite{15}. The possibility of the helicity flip for the solar neutrino, induced by torsion was predicted in Ref \cite{Ex}.

\section{Conclusions}

Over many years before, attempts to construct the theory of gravity that involves torsion have been made \cite{11} - \cite{18}, but none of the mathematical definitions have allowed for experimental torsion detection.  In this article we switch to the more general geometry, where all points of the manifold are oriented  (using the triad formalism) and the Euler angles specify the orientation of triad. We based on the geometro-hydrodynamical formalism of quantum mechanics for a spinning particle which was developed by the authors of \cite{2} - \cite{6}, where the Pauli spinor field is interpreted as defining a state of rotation   and represented in explicit geometric terms \cite{2}. The method represents the Pauli field as a motion of the fluid of an assembly of very small rotating particles continuously distributed in space and having intrinsic angular momentum.   We developed the geometry-hydrodynamical formalism and represented each element of fluid as the element of space, which is regarded as a triad having degrees of freedom of rotation. In our representation the Ricci rotation coefficients are the geometric image of torsion (\ref{New}) and
the theory predicts the existence of new torsion field with the potential equal to Ricci rotation coefficients. We can affirm that the motion of each triad vector determined by its location in the torsion field. In the context of geometry with torsion the Pauli field can be also  interpreted as a new type of physical field - torsion field. We have determined the influence of the external torsion field using the Pauli equation where the covariant derivative contains the spin connection (\ref{Hh}). We have obtained the set of hydrodynamics equations and the system of equations we constructed is comprised by  momentum balance (\ref{j3}), of the spin evolution (\ref{M3}) equations and vorticity evolution equation (\ref{T}). We show that the equation of motion (\ref{j3}) contains torsion-dependent terms that are the perfect analogy to the Lorentz force and Stern-Gerlach force field (\ref{j4}), but comprise the spin-vorticity vector $\mathbf{T}$ instead the magnetic field. We suggest this vector to be responsible for the torsion field force effect on the spinning neutral particle. This vector plays a central role in the theory and characterizes the vorticity of the space structure with the inhomogeneity of the triad field, which exerts an influence on the particle moving within it. It is important that now this new vector, which was derived by Takabayasi \cite{3}, is geometrized.   The spin evolution equation (\ref{M3}) also contains the torsion torque that causes spin precession in the external torsion field. In addition, we also have discussed the possibility of experimental detection of the torsion.

\appendix
\section{Appendix}
       Let us consider the spinning particle in the external torsion and electromagnetic field. The Dirac equation (\ref{Dirac}) for the free particle has the form
              \begin{equation}   i\hbar\frac{\partial\Psi}{\partial t}=\biggl(c\alpha_j\hat{p}^j-q\alpha_jA^j-c\frac{\hbar}{2}\alpha_j\gamma^j+\frac{\hbar}{2}\gamma_0+eA_0+mc^2\beta\biggr)\Psi, \end{equation} where the torsion has the components $\gamma_{\mu}=(\gamma_0,\gamma_j).$   Lets apply the standard procedure where for the two-component bispinor $\Psi=\begin{pmatrix}\psi \\ \varsigma\end{pmatrix}exp(-\frac{imc^2t}{\hbar})$ at low  speeds the relation is valid $\varsigma << \psi.$ After simply transformation

                    \begin{equation}   mc^2\begin{pmatrix}\psi \\ \varsigma\end{pmatrix}+i\hbar\partial_t\begin{pmatrix}\psi \\ \varsigma\end{pmatrix}=c\begin{pmatrix} 0 & \sigma_i \\ \sigma_i & 0 \end{pmatrix}\biggl(-i\hbar\hat{\nabla}^i-\frac{\hbar}{2}\gamma^i-\frac{e}{c}A^i\biggr)\begin{pmatrix}\psi \\ \varsigma\end{pmatrix}+\biggl(mc^2+q\varphi+\frac{\hbar}{2}\gamma_0\biggr)\begin{pmatrix} 1& 0 \\ 0 & -1 \end{pmatrix}\begin{pmatrix}\psi \\ \varsigma\end{pmatrix}.\end{equation} Using the relation

                      \begin{equation}    \biggl(\sigma_i\cdot(-i\hbar\hat{\nabla}^i-\frac{\hbar}{2}\gamma^i-\frac{e}{c}A^i)\biggr)^2=
                      (\delta_{ij}+i\sigma_k\varepsilon_{ijk})\biggl(-i\hbar\hat{\nabla}^i-\frac{\hbar}{2}\gamma^i-\frac{e}{c}A^i\biggr)
                      \biggl(-i\hbar\hat{\nabla}^j-\frac{\hbar}{2}\gamma^j-\frac{e}{c}A^j\biggr)
                                             \end{equation}                                      and  the low  speeds relation  $\varsigma << \psi$ we can obtain the modified Pauli equation (\ref{H}) with Hamiltonian (\ref{Hh})

                                              \begin{equation}  \label{Hh2}
                              i\hbar\frac{\partial\psi}{\partial t}=\Biggl(\frac{1}{2m}\Biggl(\hat{p}_{i}-\frac{q}{c}A_{i}-\frac{\hbar}{2}\gamma_{i}\Biggr)^2
                              +qA_{0}+\frac{\hbar}{2}\gamma_0-\mu\hat{\sigma}_{i}B^{i}_{ext}-\frac{\hbar}{2}\hat{\sigma}_{i}T^{i}\Biggr)\psi,
  \end{equation}      where the spin-vorticity vector $\mathbf{T}$ is determined by (\ref{vort}). We also use the fact that the third axis of triad $\mathbf{e}^{(3)}$ is collinear to the spin vector field and the  relation
                                          \begin{equation} \label{connection2} \gamma_{\mu}=-\frac{1}{2}\sigma_{ab}\gamma^{ab}_{\mu}=\frac{1}{2}\varepsilon_{ijk}
                                          e^{(3)}_k\gamma^{i}_{j\mu}=\frac{1}{2}\varepsilon_{ijk}
                                          e^{(3)}_k\partial_{\mu}e^{(a)}_{j}e^i_{(a)}.
                                              \end{equation}
    The semi-classical equations of motion of spinning particle in the external torsion field based on the Pauli equation (\ref{Hh2})   have the well-known form
                             \begin{equation}  i\hbar\frac{d\mathbf{\hat{x}}}{dt}=[\mathbf{\hat{x}},\hat{H}], \qquad  i\hbar\frac{d\mathbf{\hat{p}}}{dt}=[\mathbf{\hat{p}},\hat{H}], \qquad i\hbar\frac{d\mathbf{\hat{\sigma}}}{dt}=[\mathbf{\hat{\sigma}},\hat{H}].  \end{equation}
                                              The first equation gives the definition for the velocity field  \begin{equation}\label{v2}   m\mathbf{v}=\mathbf{p}-\frac{q}{m}\mathbf{A}+\frac{\hbar}{4}\sum_{a}\varepsilon^{ijk}\nabla  e_{j}^{(a)}\cdot e_{i}^{(a)}e^{(3)}_{k}, \end{equation}                          while the second and third equations (\ref{v2}) lead to the equation of motion (\ref{j3}) and spin evolution equation (\ref{M3}). In the derivation we used the commutation relations for the operator of coordinate of the particle, momentum and spin $[\hat{x}_i,\hat{p}_j]=i\hbar\delta_{ij}$, $[\hat{\sigma}_i,\hat{\sigma}_j]=2i\varepsilon_{ijk}\hat{\sigma}_k$.

 {\color{myaqua}

}}

\end{document}